# Real-Space Imaging of Guided Exciton Polaritons in Free-standing Monolayer WSe$_2$

Manuka Suriyage[1#], Hao Qin[1#], Xueqian Sun[1], Wenkai Yang[1], Shuyao Qiu[1], Qingyi Zhou[2], Zongfu Yu[2] and Yuerui Lu[1,3*]

[1]School of Engineering, College of Engineering, Computing & Cybernetics, the Australian National University, Canberra, ACT, 2601, Australia

[2]Department of Electrical and Computer Engineering, University of Wisconsin-Madison, Madison, WI 53706, USA

[3]ARC Centre of Excellence in Quantum Computation and Communication Technology ANU node, Canberra, ACT 2601, Australia

# Those two authors contribute equally to this work.

* To whom correspondence should be addressed: Yuerui Lu (yuerui.lu@anu.edu.au)

## Abstract

Monolayers of transition metal dichalcogenides (TMDCs), known for their strong excitonic states[1-3] with high binding energies[4,5] in the visible spectrum at room temperature, offer great potential for polariton-driven devices[6-10]. While polariton guided modes in bulk TMDCs have been reported[11-13] the real space experimental observation of 2D exciton-polariton guided modes in a monolayer remains challenging due to various mode cutoff conditions that arise as the TMDC layer becomes thinner, including cut-off frequency[28], mode confinement[10,29] and boundary conditions[10,29]. Here using scanning near-field optical microscopy (s-SNOM), we directly visualized the real-space propagation of these guided modes for the first time in an angstrom-thick, suspended monolayer of WSe$_2$. Through numerical simulations we have also validated that the guided mode can only exist in a monolayer WSe$_2$ when symmetric cladding conditions are closely applied. By tuning the excitation laser energy and analyzing the guided mode distribution, we observed a pronounced back-bending dispersion around the A exciton, indicating strong light-matter interactions, and confirmed the existence of the fundamental TE$_0$ exciton polariton (EP) propagation mode. The unique dispersion characteristics of these modes were further validated through theoretical modeling of the mode in free-standing monolayer WSe$_2$. Our findings provide crucial experimental evidence of guided mode EPs in atomically thin TMDCs, opening new possibilities for nanoscale photonic applications.

**Introduction**

Transition-metal dichalcogenides (TMDCs) which are van der Waals (vdW) layered materials have been studied extensively due to their exceptional electron mobility, tunable bandgaps[14] and remarkable photon absorption capabilities[15]. Light matter interactions of exciton polaritons (EPs) are widely explored for these TMDC materials in recent years by many groups using far field[16,17] and near field techniques[12,18]. Pioneer theoretical work for the demonstration of quantum well EPs[19] based on distribute Bragg reflectors made the path to realization of similar cavity mode EPs with different optical resonances coupled with the exciton resonance. Recent studies have explored various exciton complexes[20] coupled with photons, extending the investigation to unique exciton behaviors in atomically thin TMDC layers[21] and monolayers[16]. These include phenomena such as charged excitons (trions)[16] or defect exciton resonances[22], which have been used to observe distinctive cavity EP behaviors. However, cavity excitons are constrained by their radiative decay rates and small group velocities. By coupling these excitons with non-radiative waveguide modes supported by TMDCs, which exhibit large group velocities and extended propagation[13,18], they become highly advantageous for photonics applications in near IR to visible spectral range[23,24]. These waveguide mode EPs can be extremely useful for photonic applications and so far it has been used to probe optical anisotropy of thin vdW materials[10] and develop waveguided polariton lasers[25]. Guided mode EPs have been experimentally observed in multilayer TMDCs, such as WS$_2$[26,27], MoS$_2$[11], WSe$_2$[12,18], MoSe$_2$[13]. Nevertheless, observing and realizing EP guided modes in monolayers and thin layers remains challenging due to various mode cutoff conditions that arise as the TMDC layer becomes thinner, including cut-off frequency[28], mode confinement[10,29] and boundary conditions[10,29]. Despite their minimal atomic thickness, which can be just a few angstroms in the monolayer regime, these materials can support high electric currents[30,31] and exhibit strong light absorption and emission[32,33], making them highly advantageous for tuning and modifying light manipulation at the atomic level. Cavity EPs have already been experimentally realized in TMDC monolayer[16], but the potential of these monolayers to support guided mode EPs has yet to be experimentally validated.

Guided modes in plasmonic and phononic systems, such as surface plasmon polaritons[34,35] (SPPs) and surface phonon polaritons[36-38] (SPhPs), occur near the resonance where the real part of the permittivity is negative. These well-studied phenomena have been observed in various materials[34,37,38]. In contrast, although surface EPs or guided mode EPs have been predicted theoretically[39,40], they have not yet been experimentally observed in natural materials. The initial theoretical prediction of guided mode EPs in monolayer TMDCs[29] was inspired by work on SPPs, where Mikhailov and Ziegler[41] demonstrated the possibility of weakly confined TE modes in a narrow interval where graphene's permittivity becomes positive. Building on this prediction, Khurgin[29] evaluated the theoretical feasibility of a TE fundamental guided mode in monolayer TMDCs. Although a few groups have briefly discussed[10] the potential existence of guided modes in ultrathin TMDC layers, no experimental demonstration has been achieved so far.

In this work, we present the near-field observation of 2D guided mode exciton polaritons from a free-standing WSe$_2$ monolayer sample across a pre-patterned hole on a Si/SiO$_2$/Au substrate. We first discussed the fundamental physics and critical conditions necessary for the existence of these guided modes in WSe$_2$ monolayers, validating the essential requirement for symmetric cladding conditions through numerical simulations. We then present the experimental realization of these guided modes, showcasing results from s-SNOM. Initially, we highlight the distinctive features of the experimental fringes and introduce a technique for extracting the actual parameters of the propagating guided modes from the s-SNOM images. This is followed by an analysis of the experimental findings, emphasizing the unique dispersion characteristics and mode confinement observed. Finally, we demonstrate strong coupling, confirming the presence of a fundamental guided mode EP in free-standing monolayer $WSe_2$.

**Results and Discussion**

**Experimental observation of guided mode on free-standing $WSe_2$ monolayer.**

TMDC layered materials are uniaxial materials with a single out-of-plane optical axis which can support TE(ordinary) and TM(extraordinary) waveguide modes. In bulk or multilayer materials[10,11] these TE modes are governed by the in-plane dielectric constant ($\varepsilon_\perp$), and the TM modes are governed by both in-plane ($\varepsilon_\perp$) and out-of-plane ($\varepsilon_\parallel$) dielectric constants. The ability to confine these modes in a thinner sample of TMDC depends on several key conditions. $WSe_2$ monolayers which have a large effective mass and surrounded by a material with a small dielectric constant, could exhibit high peak absorption coefficients, making it possible to realize a fundamental guided mode[10,29]. The existence of a guided mode in ultrathin $WSe_2$ is highly sensitive to the refractive index balance of the surrounding cladding materials. A large refractive index mismatch between the substrate and superstrate can disrupt the $E$ field interaction with the monolayer, causing a cutoff that prevents confined mode formation. This cutoff can be avoided by closely matching the refractive indices (Supplementary Note 1; Equation S16) of the substrate and superstrate[10,29], which helps sustain mode confinement even in ultrathin layers.

To satisfy the theoretical requirements for optimal cladding conditions for ultrathin $WSe_2$, we considered a free-standing $WSe_2$ monolayer, depicted in the 2D schematic (Figure 1a). The free-standing monolayer is subjected to a s-polarized electric field (Equation S1) in our s-SNOM setup (see Supplementary Note 1 for more information). We conducted numerical simulations under varying cladding conditions to evaluate the existence of a guided mode in our monolayer sample. The simulated structure represents half of the sample shown in Figure 1a, with the AFM tip positioned centrally within the suspended monolayer $WSe_2$ sample (Figure 1b). We plot the electric field $E_y$ in the *x-z* plane, where the AFM tip's electric field confinement is modeled as a guided mode launched from the left end of the $WSe_2$ sample (indicated by the black arrow) and propagating towards the right along the x-axis with a propagation constant $\beta$ (see Supplementary Note 1). As shown in the two simulation results in Figure 1b, a guided mode is observed in the top panel, where the monolayer sample is cladded between two air layers. In contrast, in the bottom panel, where the $WSe_2$ monolayer is placed on an Au layer, the guided mode does not propagate. These simulation results confirm that a confined guided mode exists only in the suspended configuration.

Therefore, we fabricated a sample where the $WSe_2$ monolayer was transferred onto a pre-patterned hole in a $Si/SiO_2/Au$ substrate, as shown in the microscope image in Figure 1c. Using an AFM-based s-SNOM setup, we mapped out the guided modes. With the incident laser (s-polarized) focused directly on the AFM tip at an incident angle around $\alpha = 60°$ (Figure 1a) the AFM tip generates a confined electric field, launching guided modes that propagate radially through the $WSe_2$ monolayer, as illustrated in the 3D schematic in Figure 1d. As shown in the schematic in Figure 1a, when the guided modes are generated at the AFM tip and reach the circular boundary, they scatter back into free space as photons directed toward the detector along path $P_2$. These photons interfere with the photons scattered directly by the AFM tip along path $P_1$, creating an interference pattern observed in the SNOM results (Figure 1e). Despite the radial propagation of the guided modes, the SNOM results reveal parallel or slightly curved fringes. Based on this observation, we conducted a simulation to model fringe formation, considering boundary scattering conditions.

**Fringe formation numerical simulation results.**

As shown in the schematic of the SNOM experimental setup (in Supplementary Figure S2), the incident light is focused on the AFM tip at a 45° angle relative to the sample's *x-y* coordinate frame. For analysis, we define a coordinate frame ($d$-$d'$) aligned with the direction of the incident beam. When the AFM tip scans the sample in tapping mode within the circular boundary (Figure 2a) at a random point A ($d, d'$), the guided mode generated by the AFM tip radially propagates until it reaches the circular boundary. At the boundary, only two points $E_1$ and $E_2$, can scatter the guided mode into free space as photons directed toward the detector, ensuring momentum conservation along the boundary directions labeled $BE_1$ and $BE_2$ (see Supplementary Note 2).

Supplementary Figure S3b-e illustrates how scattering points on the boundary (labeled $E_1$ and $E_2$) shift based on the AFM tip's position. With the AFM tip set at a distance $d = 0.4$ μm (Figure S3b), the tip is moved along the $d'$ direction from 0 to 2 μm. As the tip moves, the boundary scattering points $E_1$ (purple region) and $E_2$ (gray region) shift, accordingly, as marked by arrowheads. When the AFM tip is positioned within 1 μm of the center (along the $d$ axis), scattering points $E_1$ and $E_2$ lie within the second and fourth quadrants of the boundary (Supplementary Figure S3b, c). However, beyond a 1 μm distance, these points transition to the first and fourth quadrants (Supplementary Figure S3d, e). These shifts adhere to the boundary scattering conditions (Supplementary Equation S19). The incident light is scattered back to the detector from the AFM tip (Path $P_1$), while additional photons scattered from guided modes at $E_1$ and $E_2$ (Path $P_2$) interfere to create the observed SNOM fringe patterns. We performed numerical analysis on this interference to determine the phase difference in the detected signal with the AFM tip position.

The phase differences between photons scattered from the boundary points ($E_1$ and $E_2$) in path $P_2$ and the photons directly scattered from the AFM tip in path $P_1$ (denotes as $\Phi_1$ and $\Phi_2$) were calculated (see Supplementary Note 2). Then by superposing these signals, we derived the total phase change of the detected signal with the AFM tip position as:

$$\Phi_{total} = arctan\left(\frac{A_1\Phi_1 + A_2\Phi_2}{A_1 + A_2}\right) \quad (1)$$

where $A_1$ and $A_2$ represent the amplitudes of scattered photons at $E_1$ and $E_2$, respectively. The results, presented in Figure 2b, show how the total phase difference varies with changes in $d'$ for constant $d$ values, confirming that the fringe patterns are primarily parallel or slightly curved. In our experimental observations, the fringe patterns lose contrast and consistency near the boundary due to complex guided-mode interferences occurring in this region. As shown in Figure 2c, when the tip is positioned far from the boundary, the incident beam is completely focused on the AFM tip which generates a tip-generated guided mode, which then scatters at the edge (path $P_2$). In this case, only the incident photons scatter back from the tip (path $P_1$) influence the experimental fringe pattern that is generated from the interference of photons collected from paths $P_1$ and $P_2$. However, when the tip is positioned closer to the boundary, as illustrated in Figure 2d, the beam simultaneously focuses on both the tip and the edge. This results in both tip-generated, and edge-generated guided modes, which interfere and create a complex electric field distribution. Consequently, the scattered photons (in path $P_1$) generate complex interference patterns (with photons collected from path $P_2$), leading to inconsistent fringe patterns near the edge.

**Dispersion analysis of the guided mode from a free standing monolayer WSe$_2$.**

Based on the above discovery through simulations and after confirming the parallel behavior of the experimental fringes the following equation was derived (see Supplementary Note 2) to determine the polariton wavelength ($\lambda_p$).

$$\lambda_p = \frac{\rho \lambda_0}{\lambda_0 - \cos(\alpha)\rho} \quad (2)$$

where $\rho$ is the measured fringe period from the SNOM images and $\lambda_0$ is the excitation wavelength. By using equation 2 we analyzed the s-SNOM imaging data collected across an energy range from 1.37 eV to 1.8 eV. Four of these experimental results, with laser energies near the exciton resonance of WSe$_2$, are shown in Figure 3a-d. We obtained the fringe period by taking a line cut from the images, considering the movement of the AFM tip from point T$_1$, near the center, to T$_2$ (2 μm away from the center) along the $d$ axis. For all the experimental measurements, we calculated the polariton wavelength using Equation 2 from the extracted $ρ$ values of each line cut profile and fitted the guided mode dispersion as illustrated in Figure 4.

**Strong coupling of the 2D guided exciton polariton on free-standing WSe$_2$ monolayer.**

Normally in cavity-mode EP measurements, a distinct separation into upper and lower polariton branches due to Rabi splitting is observed[16]. In contrast, guided-mode EPs exhibit strong coupling that introduces a non-linear dispersion characterized by back-bending due to the continuous nature of guided modes[11]. Based on our theoretical analysis (supplementary Note 1), the dispersion can be modeled as in Supplementary equation S10, and the back bending characteristics depend on the nature of the strong coupling occurs at the monolayer WSe$_2$. The fitted dispersion of the observed guided mode for the free-standing WSe$_2$ monolayer showcases a pronounce back bending confirming that the guided mode that we observed is a fundamental EP guided mode. As you can see at the exciton resonance of 1.653 eV, no fringes appeared (Figure 3b), as the energy matched the exciton transition, resulting in maximum absorption. In the lower polariton branch, spanning from 1.37 eV to 1.6 eV, polariton momentum aligned closely with the air-based dispersion. behaving similarly to an uncoupled photon mode in air due to weaker coupling away from resonance. However, near resonance at 1.642 eV, we observed a marked increase in EP momentum, as the exciton component of the polariton became more prominent and coupling was enhanced. This coupling strength near resonance leads to a high peak absorption, ( $α_{max} = 1.9$ ) and induces strong surface polarization in WSe$_2$, which dominates the wave behavior (Supplementary Note 1). This effect forces the mode to bend towards higher k values close to resonance, reducing the effective mode width $W_{eff}$ (as small as 0.2µm at the excitation energy $E$=1.646 eV) and creating confined polariton modes due to the material's 2D nature and significant surface interactions. Because of WSe$_2$'s high susceptibility, the wave becomes tightly bound to the surface, leading to strong light-matter interaction and high $k$ values. This back-bending in the polariton dispersion is a clear demonstration of strong coupling within our suspended monolayer sample (Figure 4), supporting the first observation of 2D guided mode exciton polaritons.

**Conclusion**

In this study we have experimentally demonstrated the existence of 2D EP guided modes in free-standing monolayer TMDC materials. Our work confirms the theoretical predictions and discusses more on the essential conditions and behaviors of these guided modes with comparison to cavity and bulk exciton polaritons that have been investigated extensively in the recent past. We conducted a comprehensive numerical analysis of fringe formation in the s-SNOM system, effectively linking the observed fringe periodicity to the intrinsic polariton wavelength. This correlation is key to understanding the scattering behavior of a suspended monolayer in an s-SNOM setup, offering valuable insights for guiding future experimental investigations. Through the experimental findings, we also highlight the unique dispersion relations, observing a pronounced back-bending near the A exciton resonance, which reveals strong light-matter interactions specific to monolayer WSe$_2$. The successful realization of EP guided modes in such atomically thin material underscores the potential of monolayer TMDCs for advanced photonic applications, particularly in the near infrared to visible frequency range. This breakthrough paves the way for further exploration of light-matter interactions in 2D materials and could lead to the development of novel photonic devices that leverage the extraordinary properties of exciton polaritons in atomically thin systems.

## Methods

### Numerical Simulations

We use COMSOL Multiphysics software to simulate the electric field distribution of the guided mode EPs in thick and ultra-thin $WSe_2$ samples which are subjected to different cladding conditions. We performed simulations for different substrate and superstrate combinations to validate the cladding effect towards the confinement of exciton polaritons in thick and thin samples. For ultra-thin samples we performed simplified 2D simulations to investigate the guided mode exciton polaritons in monolayer $WSe_2$ clad between two air layers to justify and analyze the experimental results obtained. More simulation details can be found in Supplementary Note 4.

### Materials and Fabrication

Our free-standing $WSe_2$ monolayer sample was mechanically exfoliated from bulk crystals and subsequently dry-transferred onto an $Si/SiO_2/Cr/Au$ (Cr/Au – 3nm/70nm) substrate featuring circular hole patterns. A pre-etched circular hole with a diameter of 5 μm and a depth of 3 μm was selected from the substrate to achieve the suspended structure. The number of layers was characterized by the contrast of optical microscope and room-temperature photoluminescence measurement.

### s-SNOM measurements

Room-temperature real-space imaging of exciton polaritonic patterns was achieved using a commercially available scattering-type scanning near-field optical microscopy (s-SNOM) system (Neaspec). This system utilized a tapping-mode atomic force microscope (AFM) equipped with Pt/Ir-coated silicon tips (Nano World). The tapping resonance frequency was approximately 285 kHz, and the tapping amplitude was set to around 70 nm. A Maitai femtosecond pulse laser with an 80 MHz repetition rate and a tunable range from 1.192 eV to 1.796 eV was optically guided into the s-SNOM. The power and polarization were controlled using a combination of a half-wave plate (WPH05M-780, Thorlabs) and a polarizing beam splitter (PBS122, Thorlabs), ensuring that only s-polarized laser light was directed to the near-field setup. The near-field signal was detected by a pseudo-heterodyne interferometric detection module, which decoupled the scattering amplitude (s) and phase (ψ) of the near-field signal. The scattering amplitude (s) signal was demodulated to the third harmonic ($S_3$) of the AFM tip's tapping frequency to present the characteristics of the exciton polariton with background signal subtracted.

**Data Availability**

The data that supports the findings of this research are available from the corresponding authors upon reasonable request.

**Author Contributions**

Y.L. conceived and supervised the project; X.S and W.Y prepared the suspended monolayer samples; H.Q and S.Q carried out all the measurements; M.S, X.S and Y.L performed the theoretical analysis; M.S, X.S, H.Q analyzed the data, M.S performed the analytical calculations, M.S, W.Y and Q.Z carried out all the simulations, M.S and H.Q drafted the manuscript and all authors contributed to the manuscript.

**Supplementary Information**

All additional data and supplementary information and methods are presented in the supplementary information file available online. The figures and information in the supplementary information have been cited at appropriate places in this manuscript.

**Acknowledgements**

The authors acknowledge funding support from ANU PhD student scholarship, Australian Research Council (ARC; numbers DP220102219, DP180103238, LE200100032) and ARC Centre of Excellence in Quantum Computation and Communication Technology (CE170100012). Qingyi Zhou and Zongfu Yu acknowledge support from National Science Foundation's QLCI-CI: Hybrid Quantum Architectures and Networks.

**Competing financial interests**

The authors declare that they declare that they have no competing financial interests.

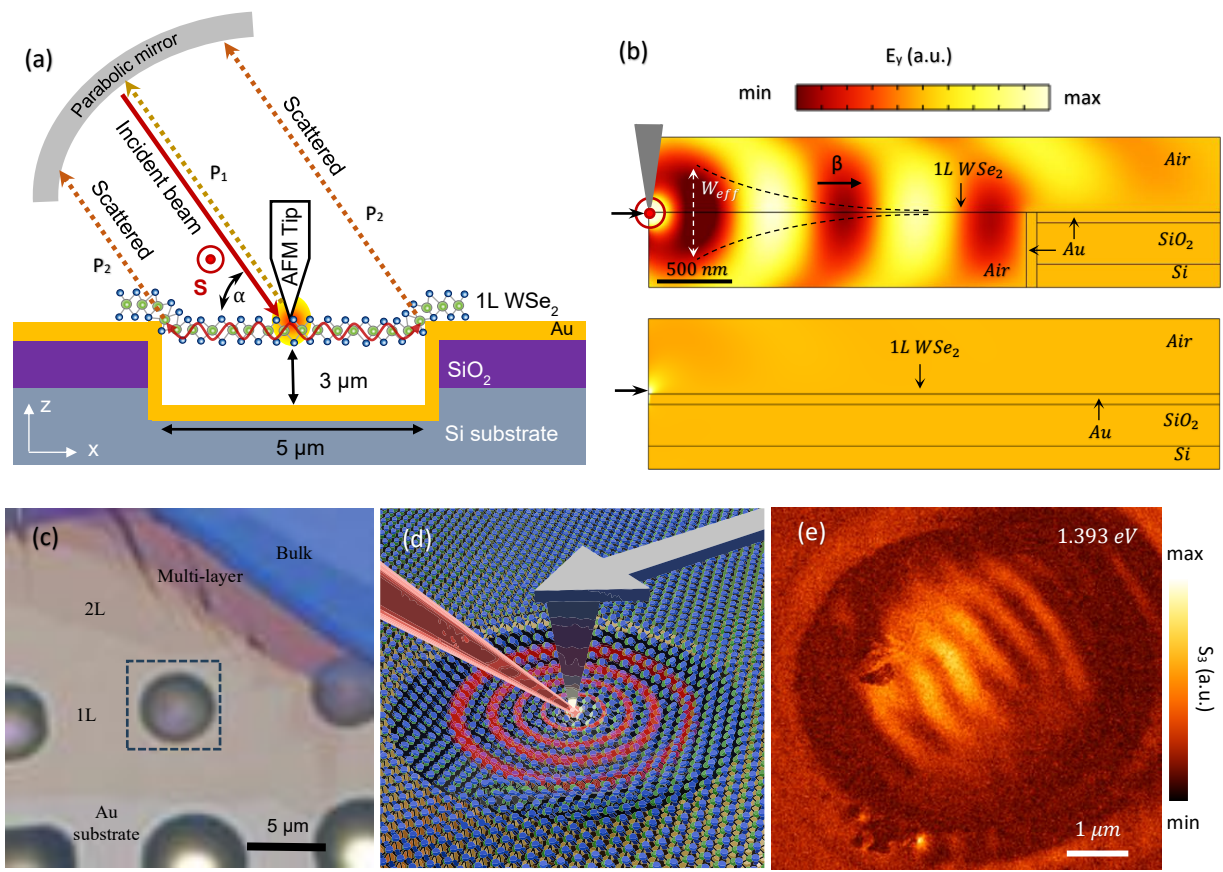

**Figure 1 | Experimental observation of guided mode on free-standing WSe₂ monolayer.**
**a,** 2D schematic of a free-standing WSe₂ monolayer over a pre-patterned Si/SiO₂/Au substrate. The hole has a depth of 3 μm and a diameter of 5 μm. $\alpha \approx 60°$, $\alpha$ is the angle between the s-polarized incident beam (red solid arrow) and sample surface. P₁ is the path of tip scattered photons (yellow dotted arrow) and P₂ is the path of edge scattered photons (orange dotted arrow) **b,** Simulated 2D guided mode in half of the free-standing monolayer sample, illustrating the tip-generated mode confinement and propagation when the tip is positioned at the center of the free-standing monolayer region, when the sample is excited by photons at energy $E = 1.393\ eV$. Guided mode propagates in the along the x axis with a $\beta$ propagation constant and the $W_{eff}$ is the effective mode width of the mode. **c,** Optical microscope image of the sample, highlighting the Au regions, monolayer, bilayer, and bulk areas. **d,** 3D schematic of the top view of the experimental setup to visualize the 2D guided mode on a free-standing WSe₂ monolayer. **e,** Experimental near-field image of the 2D waveguide mode from a free-standing monolayer WSe₂ at the illumination Energy $E = 1.393\ eV$.

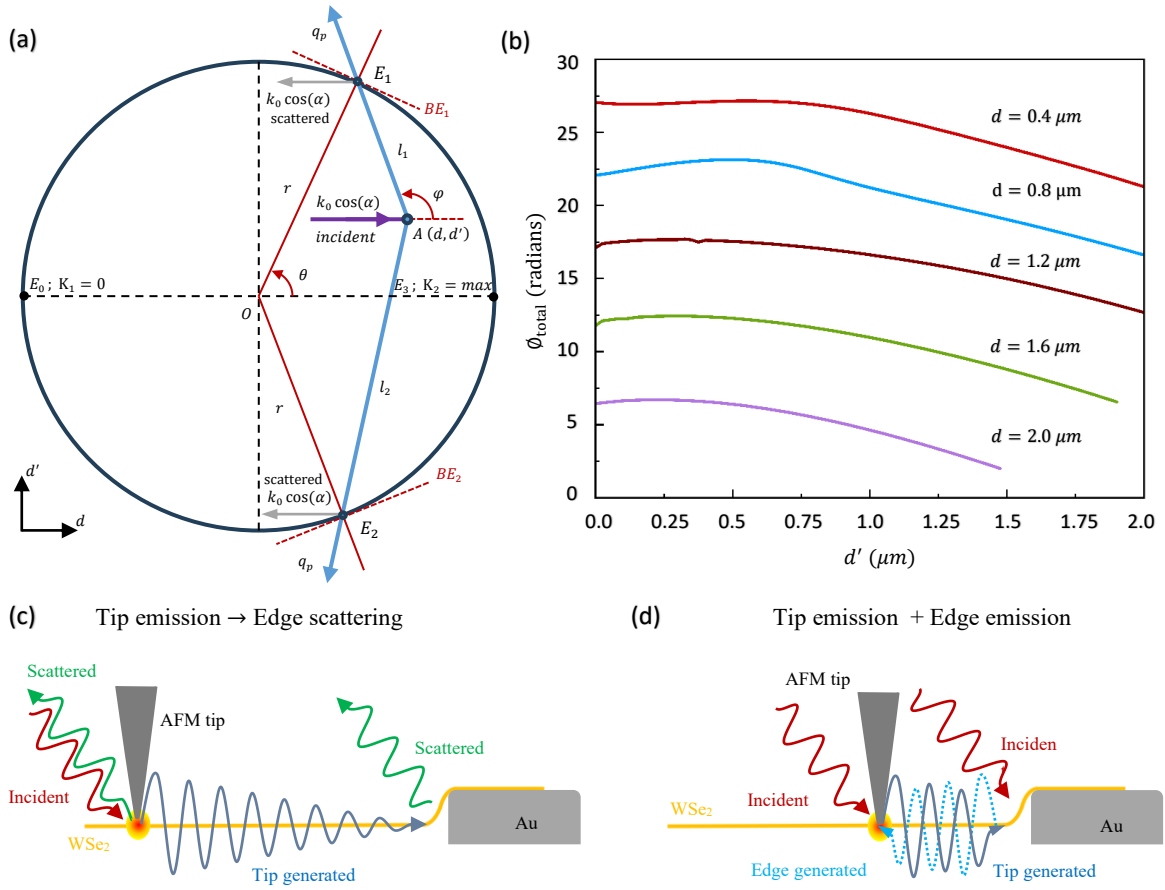

**Figure 2 | Fringe formation numerical simulation results. a,** 2D Schematic of the guided mode with a wave vector $q_p$ scattering at boundary points $E_1$ and $E_2$ along the circular edge, producing free-space photons ($k_0$) directed towards the detector. $r = 2.5 \mu m$, where r is the radius of the hole. $l_1$ and $l_2$ are the propagation distances of the guided modes propagating from the AFM tip position to $E_1$ and $E_2$ scattering points. $K_1$ and $K_2$ are the guided mode edge scattering efficiencies for the two scattering points. $E_0$ and $E_3$ are the two points in the boundary with minimum and maximum boundary scattering efficiency. $\theta$ is the angle between $OE_1$ and the $d$ axis. $\varphi$ represents the angle of $q_p$ w.r.t the $d$ axis. $d$-$d'$ coordinate frame is 45° rotated relative to the $x$-$y$ sample coordinate frame to align the actual incident laser direction along the $d$ axis. $BE_1$ and $BE_2$ are the scattering boundary edge directions at scattering points $E_1$ and $E_2$. **b,** Numerical simulation result of the total phase difference in the SNOM-detected signal for various AFM tip positions at fixed $d$ coordinates: $d = 0.4, 0.8, 1.2, 1.6$ and $2 \mu m$. **c, d,** Schematic illustrations of polariton generation and scattering conditions: (c) when the AFM tip is positioned near the center of the circular boundary, and (d) when the tip is closer to the boundary edge.

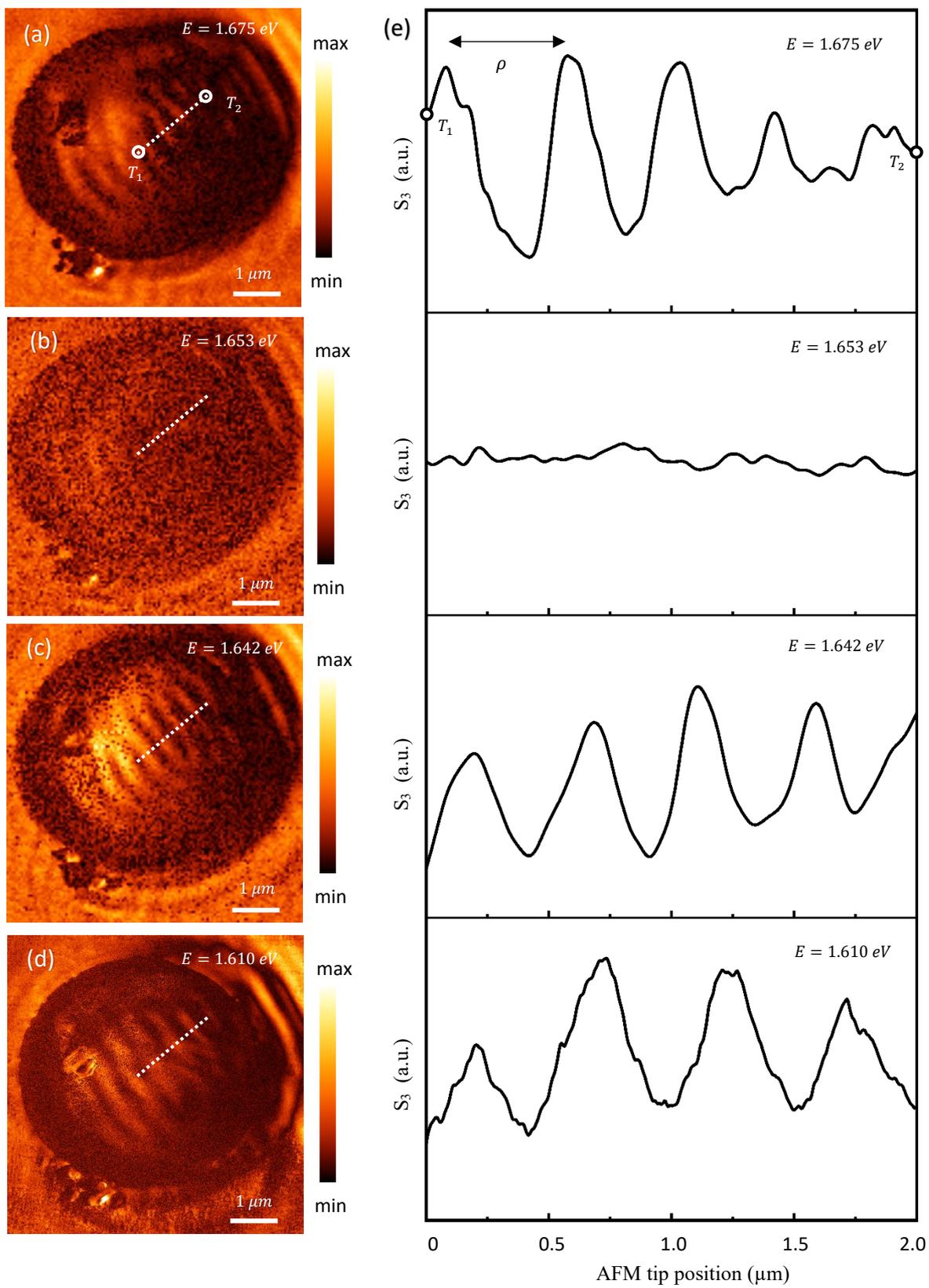

**Figure 3 | Dispersion analysis of the guided mode from a free standing monolayer WSe$_2$.**
**a-d,** Experimental near-field images of guided modes in suspended monolayer WSe$_2$ at the excitation Energies of $E = 1.675, 1.653, 1.642$ and $1.610\ eV$. Color bar represents the intensity of the s-SNOM amplitude signal $S_3$. $T_1$ and $T_2$ are the AFM tip positions at the start and the end of the line cut profile. **e,** Line cut profile data for the near field images a-d parallel to the $d$ axis and closer to $d' = 0$. $\rho$ is the measured experimental fringe period.

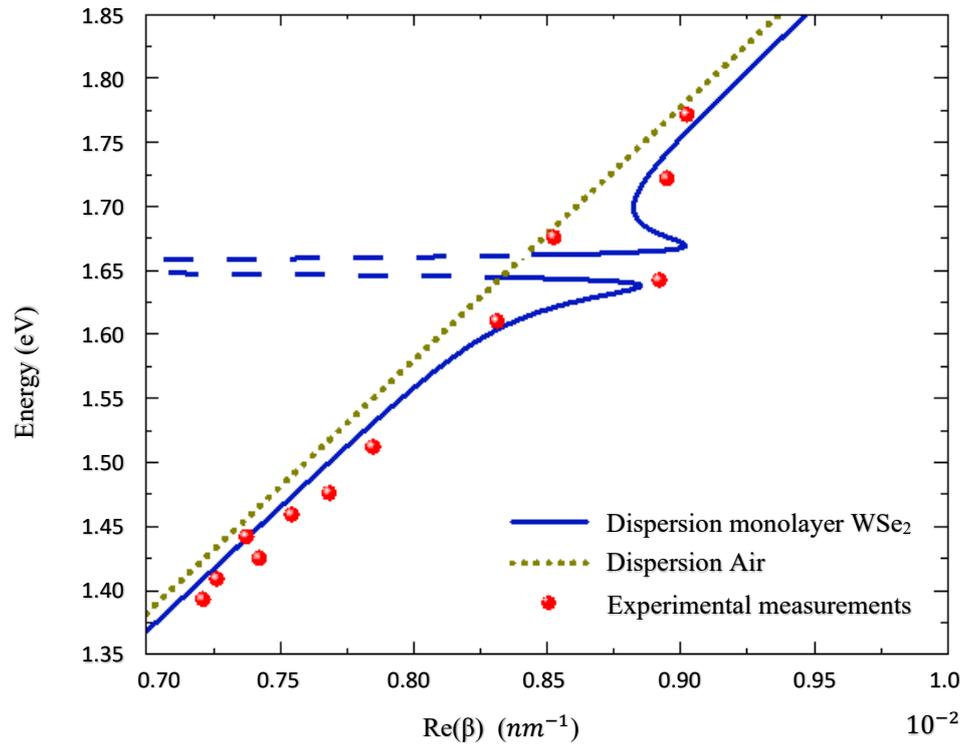

**Figure 4 | Strong coupling of the 2D guided exciton polariton on free-standing WSe$_2$ monolayer.** Simulated dispersion relation of a 2D guided exciton polaritons alongside experimentally measured wave vectors obtained from near-field imaging results.